\documentclass[letterpaper, 10 pt, conference]{ieeeconf}  

\IEEEoverridecommandlockouts                              

\usepackage[utf8]{inputenc}
\usepackage{amsmath}
\usepackage{graphicx}
\usepackage{amsfonts}
\usepackage{xcolor}
\usepackage{amsmath,amssymb,amsfonts}
\usepackage{subfig}

\IEEEoverridecommandlockouts 

\title{Reinforcement Learning-based Optimal Control and Software Rejuvenation for Safe and Efficient UAV Navigation
}
\author{Angela Chen, Konstantinos Mitsopoulos, and Raffaele Romagnoli
\thanks{A. Chen and R. Romagnoli are  with the Department of Electrical and Computer Engineering, Carnegie Mellon University, 5000 Forbes Ave, Pittsburgh, PA 15213, USA:
        {\tt\small  xinyuc2@andrew.cmu.edu}
        {\tt\small rromagno@andrew.cmu.edu}}%
\thanks{Konstantinos Mitsopoulos is with the Institute for Human and Machine Cognition, 40 South Alcaniz St, Pensacola, FL 32502, USA:
        {\tt\small  kmitsopoulos@ihmc.org }}%
}

\begin{document}

\maketitle
\begin{abstract}
Unmanned autonomous vehicles (UAVs) rely on effective path planning and tracking control to accomplish complex tasks in various domains. Reinforcement Learning (RL) methods are becoming increasingly popular in control applications \cite{bertsekas2019reinforcement}, as they can learn from data and deal with unmodelled dynamics. Cyber-physical systems (CPSs), such as UAVs, integrate sensing, network communication, control, and computation to solve challenging problems. In this context, Software Rejuvenation (SR) is a protection mechanism that refreshes the control software to mitigate cyber-attacks, but it can affect the tracking controller's performance due to discrepancies between the control software and the physical system state. Traditional approaches to mitigate this effect are conservative, hindering the overall system performance. In this paper, we propose a novel approach that incorporates Deep Reinforcement Learning (Deep RL) into SR to design a safe and high-performing tracking controller. Our approach optimizes safety and performance, and we demonstrate its effectiveness during UAV simulations. We compare our approach with traditional methods and show that it improves the system's performance while maintaining safety constraints.
\end{abstract}

\section{Introduction}
Path planning and tracking control are key elements for unmanned autonomous vehicles (UAVs). Reinforcement Learning \cite{sutton2018reinforcement}is gaining more attention in control applications \cite{levine2018reinforcement} since it is able to deal with unmodelled dynamics by learning them from data. UAVs are applications of cyber-physical systems (CPSs) that integrate sensing, network communication, control, and computational methods to solve complex applications in applications such as transportation, healthcare, power supply, etc \cite{platzer2018logical}.  
In general, the controller design considers only the physical dynamics of a CPS because it is assumed that the inertia of the physical system is slower than any operation performed in the cyber part. Due to the complexity of a CPS, that assumption is getting more and more unrealistic, particularly when solutions to protect the CPS from cyber-attacks are implemented \cite{AM17,DPF+19,HLLL17}. This is the case of software rejuvenation (SR) \cite{arroyo2019yolo, abdi2018preserving} which is a mechanism of protection that refreshes the run-time control software in order to mitigate the possible negative effects of a cyber-attack on it. This mechanism of protection imposes constraints to be satisfied, for example in \cite{romagnoli2023software}, the trajectory setpoints must be updated only under specific conditions that involve time and system dynamics. Despite its effectiveness, in terms of safety and mission liveness \cite{romagnoli2019safety}, the overall control performance can be very poor in terms of trajectory tracking. One of the main issues is that there is a discrepancy between the state of the control software and the actual state of the system. In fact, at each software refresh a previous uncorrupted image of the control software is loaded. This discrepancy becomes more evident in the case of the controller making use of the state estimation. Modeling this aspect into the physical system dynamics in order to develop a tracking controller that mitigates this effect can be very challenging. 
In the SR framework, the trajectory tracking controller generates a sequence of setpoints that takes into account safety accordingly to Lyapunov's theory \cite{romagnoli2019design}. In real applications, the effects of the software rejuvenation on the state estimation error make it difficult to be modeled and find the optimal trajectory tracking algorithm that can improve the overall system performance. In fact, the proposed solutions are quite conservative which is good from the safety viewpoint, but this can be very limiting in terms of the application viewpoint. 

In this paper, we involve Deep RL in the SR problem for the design of a safe tracking controller that also considers the system's performance during the mission. Our objective is to show the applicability and effectiveness of Deep RL in this context tracing the way of future research directions that combines control theory and Deep RL for safe-critical applications. 

Deep RL algorithms learn optimal control policies by iteratively optimizing a reward function that measures the success of the control policy. While this approach have been successfully applied to many control problems, in this work we do not intend to replace traditional control methods. Rather, the objective is to integrate it into existing control frameworks to improve their performance and safety.
In this paper, we apply it to the SR problem in control theory and demonstrate its potential to enhance the safety and performance of UAVs. Specifcially, we show that our apporach mitigates the effect of SR on the tracking controller's performance compared to traditional approaches. Our work contributes to the growing body of research that combines control theory and Deep RL to address critical safety issues in cyber-physical systems.



\section{Preliminaries}
Let us consider a positive definite matrix $M>0$ with $M\in \mathbb{R}^{n}$, and a vector $v\in \mathbb{R}^n$, the norm of $v$ w.r.t $P$, $P-$norm is
\begin{equation}
    \Vert v \Vert_P = \sqrt{v^TPv}.
\end{equation}
The ellipsoid of size $\rho$ centered in $c \in \mathbb{R}^n$ 
\[
\mathcal{E}(\rho,c) = \left\lbrace v \in \mathbb{R}^n\; |\; \Vert v - c\Vert_P^2 \leq \rho \right\rbrace
\]

A linear time-invariant (LTI) continuous-time system is described by
\begin{equation}\label{eqn:lti}
    \dot x = Ax+Bu
\end{equation}
where $x\in \mathbb{R}^n$ represents the state of the system, $u\in \mathbb{R}^p$ is the input vector,  the matrix $A \in \mathbb{R}^{n \times n}$, and the matrix $B \in \mathbb{R}^{n \times p}$. The output vector is $y=Cx$ with $y\in \mathbb{R}^q$ and $C \in \mathbb{R}^{q \times n}$. Let us consider a state feedback controller $u=-Kx$ that defines the closed-loop system 
\begin{equation}\label{eqn:lti_cl}
    \dot x = (A-BK)x
\end{equation}
where matrix $A-BK$ is Hurwitz. Since the controlled system is asymptotically stable there exists a positive definite matrix $P>0$ with $P \in \mathbb{R}^{n \times n}$ that satisfies the Lyapunov equation
\begin{equation}
    (A-BK)^TP + P(A-BK) = -Q
\end{equation}
where $Q \in \mathbb{R}^{n\times n}$ and $Q>0$. 
The ellipsoid centered at the origin $\mathcal{E}(\rho,0)$ is a Lyapunov level set which is positively invariant. Moving the system to another equilibrium point (or setpoint) $x_{sp}$ the new control law is $u=-K(x-x_{sp})$, then the closed-loop system can be rewritten as
\begin{equation}\label{eqn:lti_cl_sp}
    \dot x = (A-BK)(x-x_{sp}).
\end{equation}
The Lyapunov analysis remains the same except for the origin of the system which is translated as the ellipsoid $\mathcal{E}(\rho, x_{sp})$. In case the state of the system is not measurable, and only the measurements $y$ are available, a state estimation $\hat{x} \in \mathbb{R}^n$ can be used to close the loop. If the system is observable, thanks to the separation principle it is possible to design a deterministic observer 
\begin{equation}\label{eqn:observer}
    \dot{\hat{x}} = A \hat{x} + Bu+L\left(C\hat{x}-y\right).
\end{equation}
By defining the estimation error $e\triangleq x-\hat{x}$, and substituting $y$ with $Cx$, the dynamics of the estimation error is
\begin{equation}
  \dot e = (A-LC)e.  
\end{equation}
Thanks to the observability property of the system it is possible to design $L$ that makes the matrix $(A-LC)$ Hurwitz. The new control input is now 
\begin{equation}\label{control_est}
    u= -K(\hat{x}(t)-x_{sp})
\end{equation}
\section{Software Rejuvenation}
\begin{figure}[!htb]
    \centering
    \includegraphics[width=0.49\textwidth]{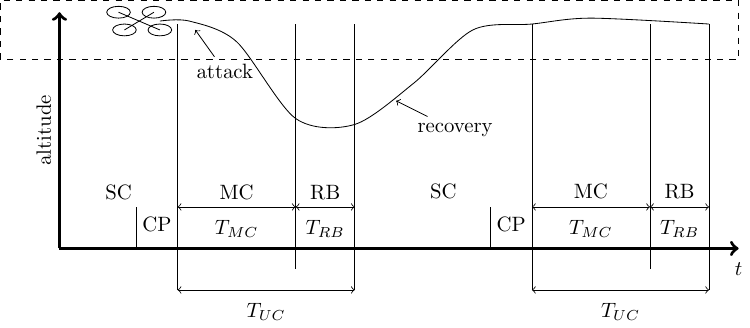}
    \caption{Software Rejuvenation timeline.}
    \label{fig:time_line}
\end{figure}

\begin{figure}[!htb]
    \centering
    \includegraphics[width=0.49\textwidth]{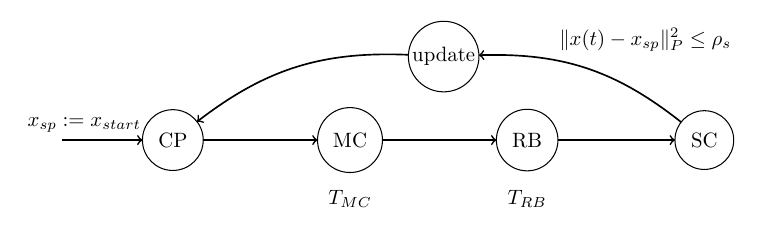}
    \caption{SR mode-transition graph. Unlabeled transitions occur immediately after the operations for the preceding mode are completed, or when the time indicated for the mode has elapsed.}
    \label{fig:ES_switching}
\end{figure}

Fig. \ref{fig:time_line} describes the SR approach over time. At the beginning of the mission the drone is in secure control ($SC$) mode which means that the control software is not vulnerable to attacks (e.g. not connected to the communication network). Before switching to mission control ($MC$) mode an image of the run-time software can be saved in a protected memory location, checkpoint ($CP$) since the system is assumed to be clean. 
During $MC$ the drone can communicate through the communication network and then be vulnerable to cyber-attacks. To avoid possible catastrophic consequences of a worst-case attack, a protected timer triggers the software refresh before it is too late for preventing any irreversible damage to the system. The amount of time the system is in $MC$ mode is indicated by $T_{MC}$. 
During software refresh, the saved clean image of the run-time software is rolled back ($RB$), and the time needed for this operation is indicated with $T_{RB}$. During $RB$, the control input is kept constant and equal to the last provided. The total time the system is under unknown control is $T_{UC}=T_{MC}+T_{RB}$. Fig. \eqref{fig:ES_switching} shows the mode-switching graph and it offers more details in particular for the recovery and setpoint update.  
For the moment we consider that $x(t)$ is available and it is used to compute the control law, hence the time spent by the system in SC mode depends on the following condition
\begin{equation}\label{setpoint_condition}
    \Vert x(t) - x_{sp} \Vert_P^2 \leq \rho_s
\end{equation}
that determines that the system has been fully recovered, and $0<\rho_s<1$. Since after $RB$, the software is clean and the system is in $SC$ mode, and $x_{sp}$ is the same saved during the previous $CP$, all the information used in \eqref{setpoint_condition} is not corrupted. If there is no attack, the time spent in $SC$ mode can last only the period to check \eqref{setpoint_condition}.

\subsection{Safety and Setpoint Update}\label{safety SR}
\begin{figure}[!htb]
    \centering
    \includegraphics[scale=0.5]{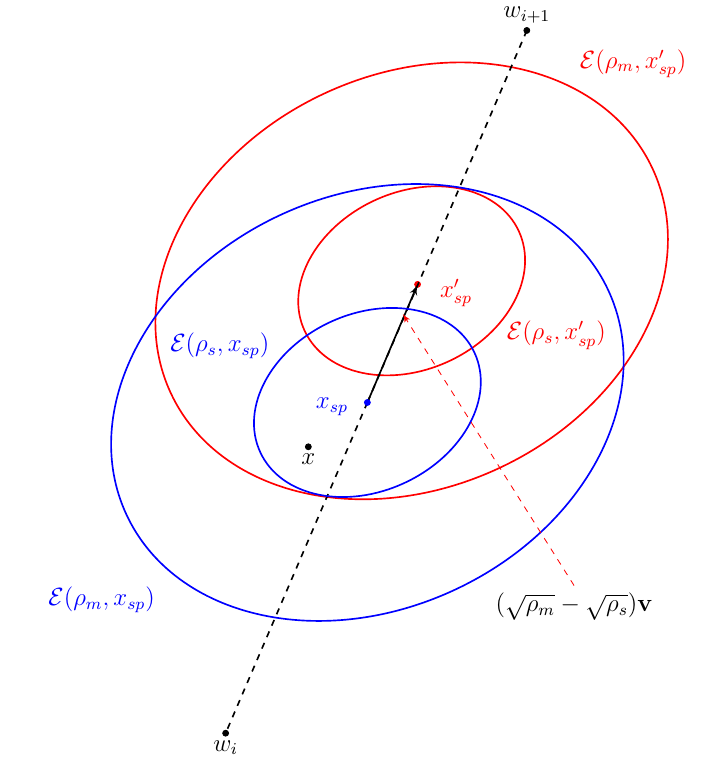}
    \caption{Safe setpoint transition scheme.}
    \label{fig:tracking}
\end{figure}

 For a given setpoint $x_{sp}$, the safety set is provided by the ellipsoid $\mathcal{E}(1,x_{sp})$ which is an invariant set for the controlled system \eqref{eqn:lti_cl_sp}. Considering a $\rho_m$ such that $0<\rho_s<\rho_m<1$, we compute $T_{MC}$ as the time that $\forall x(t) \in \mathcal{E}(\rho_m,x_{sp})$, $x(t)$ is always recoverable into $\mathcal{E}(\rho_s,x_{sp})$ \cite{romagnoli2019design}. 
For the trajectory tracking, we assume that the setpoints $x_{sp}$ are generated along the line that joins two waypoints $w_i$, and $w_{i+1}$. The safety condition for the setpoint transition is 
\begin{equation}
    \Vert x(t)-x_{sp}\Vert_P^2\leq \rho_m \Rightarrow \Vert x(t)-x_{sp}'\Vert_P^2\leq \rho_m
\end{equation}
where $x_{sp}'$ is the new setpoint \cite{romagnoli2019safety}.  Fig.\ref{fig:tracking} shows the Assuming that the state is available, the above condition is verified if $x_{sp}$ is updated as 
\begin{equation}\label{update_formula}
    x_{sp}'=x_{sp}+(\sqrt{\rho_m}-\sqrt{\rho_s})\mathbf{v}
\end{equation}
where $\mathbf{v}$ is the unitary vector along the trajectory. 
\subsection{State Estimation}
In a real application, only the state estimation $\hat{x}$ is available. This information is stored in the run-time software and during $RB$ the state estimation is computed starting from the initial conditions saved during $CP$. Since $\hat{x}$ is used to compute $u$ \eqref{control_est}, after each $RB$ there is the effect of the estimation error $e$ that may increase after each SR cycle making the system unstable. Moreover, $\hat{x}$ is now used for evaluating \eqref{setpoint_condition}, and large estimation can make the SR scheme switch when the system cannot be exposed to possible attacks. In this situation, the safety conditions for $T_{MC}$ and setpoint generation, can be the same by just replacing $x(t)$ with $\hat{x}(t)$ if $T_{est}$ is introduced \cite{romagnoli2020robust}. $T_{est}$ is the minimum time for the system to be in $SC$ mode after software refresh in order to reduce the estimation error to keep the system stable and safe against attacks. 

 \subsection{Problem Statement}
In this paper, we consider a UAV whose nonlinear dynamics can be reduced into the form of \eqref{eqn:lti} and stabilized around a setpoint $x_{sp}$ by a linear controller \eqref{eqn:lti_cl}. We also assume that the state $x$ is not directly accessible and a state observer \eqref{eqn:observer} is needed to compute $u$ as in \eqref{control_est} and for evaluating the SR condition \eqref{setpoint_condition}.
We assume that $\mathcal{E}(\rho_s, x_{sp})$, $\mathcal{E}(\rho_m, x_{sp})$, $T_{MC}$, and $T_{est}$ given and they ensure safety against cyber-attacks \cite{romagnoli2023software}.

In this paper, we are interested in the improvement of the performance of the system protected via SR when the system is not under attack. Specifically, we redesign \eqref{update_formula} as
\begin{equation}\label{update_NN}
    x_{sp}'= x_{sp} + \alpha\mathbf{v}
\end{equation}
with
\begin{equation}
   \alpha = f(\hat x, x_{sp}; \mathbf{\theta}) 
\end{equation}
where $f$ is a neural network with parameters $\mathbf{\theta}$ optimized with RL. This formulation aligns with reference governor (RG) \cite{taylor2021data, li2021safe} and explicit reference governor (ERG) \cite{liu2019model} frameworks. The safe-trajectory controller for SR can be regarded as an ERG, albeit with distinct operating conditions.
With our method we aim to show that the effects of the SR scheme can be captured by a learning technique which it would be difficult to model with the traditional control tools as shown in \cite{romagnoli2023software}. The goal is to demonstrate that the efficacy of RL approach improves performance in terms of reducing the time of the mission, while the safety conditions are satisfied. 
Finally, we also consider the presence of noise in the measurements.



\section{Learning Setpoint Generation}
We consider a task where a UAV is required to navigate from a starting location A to a goal location B within a bounded 3D space free of obstacles. The UAV is controlled by \eqref{control_est}, and we assume that an RL agent must effectively learn to modulate the displacement of a setpoint at discrete timesteps, based on input information related to the UAV's state. To accomplish this, the agent must select an appropriate value for the parameter $\alpha$, which determines the magnitude of the displacement modulation. Between two consecutive decision-making points in simulation, disturbances SR affect the UAV, including state estimation errors that depend on the agent's choice of $\alpha$. Generally, a higher value of $\alpha$ leads to a greater degree of disturbance experienced by the UAV. The main objective of the learning agent is to identify an optimal value of $\alpha$ that can modulate the UAV's displacement in a way that respects the safety constraints - discussed in Section \ref{safety SR} - while simultaneously improving the speed of the UAV.

In contrast to \eqref{update_formula}, which employs a more conservative approach that prioritizes safety but overlooks performance, our proposed approach seeks to optimize both safety and performance. Specifically, by using a learning agent that can dynamically adjust the value of $\alpha$ in response to the UAV's state, we can achieve a better balance between safety and performance, leading to improved task outcomes.

\subsection{Reinforcement Learning}
We formulate this task as a Markov Decision Process \cite{MDPPuterman} (MDP) which is defined as a tuple   $\left\langle\mathcal{S}, \mathcal{A}, T, \mathcal{R}, \gamma\right\rangle$ where:

\begin{itemize}

    \item $\mathcal{S}$ denotes the continuous state space; in our case the state at decision timestep $k$ is $s_k=\hat{x} - x_{sp}$,
    \item $\mathcal{A}$ the continuous action space; in our case the action at timestep $k$ is $a_k = \alpha \in [0,1]$,
    \item $\mathcal{T}$ is the transition probability for arriving into state $s_{k+1}$ when executing action $a_k$ from state $s_k$,
    \item $\mathcal{R}$ is the reward function that defines the reward received by the agent for transitioning from state $s_k$ to state $s_{k+1}$ when taking action $a_k$,
    \item and $\gamma$ is the discount factor that determines the importance of future rewards relative to immediate rewards. In this case, it can be used to model the trade-off between short-term and long-term objectives.
    
\end{itemize}

The objective of the agent is to maximize the expected return $G_{k}=\sum_{l=0}^{\infty} \gamma^{l} r_{k+l+1}$ from each state $s_{k}$, where $r_k$ denotes a specific instance of the reward function, obtained at evaluating it at a specific state-action pair. The reward function in our case is defined in section \ref{sec:sim_env}. 

A solution to an MDP is obtained by finding an optimal policy $\pi\left(. \mid s_{k}\right)$, that maps a state $s_{k}$ to a distribution over possible actions that lead the agent to higher sums of rewards. The probability of performing action $a_{k}$ in state $s_{k}$ is denoted by $a_k \sim \pi\left(a \mid s_{k}\right)$.

One way to obtain an optimal policy is to use value-based RL methods. The action value $Q^{\pi}(s, a)=\mathbb{E}\left[G_{k} \mid s_{k}=s, a\right]$ is the expected return for selecting action $a$ in state $s$ and following policy $\pi$. The optimal value function $Q^{*}(s, a)= \max _{\pi} Q^{\pi}(s, a)$ gives the maximum action value for state $s$ and action $a$ achievable by any policy. Similarly, the value of state $s$ under policy $\pi$ is defined as $V^{\pi}(k)=$ $\mathbb{E}\left[G_{\mathrm{t}} \mid s_{k}=s\right]$ and is simply the expected return for following policy $\pi$ from state $s$. The optimal state value function is given by $V^{*}(s)= \max _{a\in \mathcal{A}} Q^{{\pi^*}}(s, a)$.
Value functions can be used to define a policy (e.g. $\epsilon$-greedy). RL methods that estimate value functions are usually called \textit{critic methods}.

 In many real-world scenarios, the state and action spaces of an MDP are so large that it is impractical to enumerate all possible combinations. For an agent to learn a successful policy it is necessary to be able to estimate value functions of unseen states. The action value function could be represented using a function approximator, such as a neural network. Let $Q(s, a ; \theta)$ be an approximate action-value function with parameters $\theta$. The updates to $\theta$ can be derived from a variety of reinforcement learning algorithms which aim to directly approximate the optimal action value function: $Q^{*}(s, a) \approx Q(s, a ; \theta)$.

In contrast to value-based methods, policy-based model-free methods directly parameterize the policy $\pi(a \mid s ; \theta)$ and update the parameters $\theta$ by performing, typically approximate, gradient ascent on $\mathbb{E}\left[G_{k}\right].$ One example of such a method is the REINFORCE family of algorithms \cite{williams1992simple} which updates the policy parameters $\theta$ in the direction $\nabla_{\theta} \log \pi\left(a_{k} \mid s_{k} ; \theta\right) G_{k}$. Such types of methods are called \textit{actor methods}. As discussed, we can introduce an estimation of the return in the form of a critic which results in Actor-Critic methods.


In this work we employ the Soft Actor Critic \cite{SAChaarnoja2018} (SAC) algorithm to demonstrate the importance of learning methods in improving performance and ensuring safety in control applications. SAC  is an entropy-regularized RL method that changes the RL problem (i.e obtain an optimal policy $\pi^*$) to:
\begin{equation}
\begin{split}
\pi^*=\arg \max _\pi {\mathbb{E}}\left[\sum_{k=0}^{K} \gamma^k\left(R\left(s_k, a_k\right)
+\beta H\left(\pi\left(\cdot \mid s_k\right)\right)\right)\right]
\end{split}
\end{equation}
where the temperature parameter $\beta$ controls the stochasticity of the optimal policy as it determines the relative importance of the entropy $H$ of the policy term against the reward.
SAC incorporates a modified action and state value functions that offer the agent a bonus proportionate to the policy's entropy. This approach renders policies optimized for maximum entropy (\cite{jaynes1957information, ziebart2008maximum}) more robust, allowing for a greater ability to respond successfully to unexpected perturbations during testing. Additionally, optimizing for maximum entropy during training can improve both the algorithm's robustness to hyperparameters and its sample efficiency, making SAC a useful tool for control problems \cite{haarnoja2018acquiring}.

\section{Experimental setup and Simulation Results}
\subsection{Simulation Environment}
\label{sec:sim_env}

To simulate the interaction between the RL agent and the UAV system, we developed a customized OpenAI gym environment. The environment models the nonlinear dynamics of the UAV system \cite{romagnoli2023software}, along with the effects of the software rejuvenation and recovery periods. The state estimation $\hat x(t)$, computed as \eqref{eqn:observer}, is evaluated after each cycle of the SR scheme of Fig. \ref{fig:ES_switching}, with $T_{MC}=200$ ms, $T_{RB}=10$ ms, and $T_{est}=1.7$ s. Those numbers have been computed accordingly to \cite{romagnoli2023software}. The total time needed for one cycle of SR is at least $1.910$s. At approximately every 2s interval, the RL agent receives the current state $s_k$ of the system and selects an action $a_k=\alpha$, indicating its displacement from the current location as depicted in \ref{fig:tracking}. Based on \eqref{update_formula}, the $\alpha$ value is bounded as $0 \leq \alpha \leq \sqrt{\rho_m}-\sqrt{\rho_s}$ for safety considerations, and we set the size of the outer ellipsoid $\rho_m=0.01$ and the inner ellipsoid $\rho_s=0.0012$. In this experiment, the drone starts from (1,1,1) and stops when it reaches (5,5,5).


\textbf{Reward Function:} The reward function considers the effect of the action, generated by the agent, to the SR period and how far from the goal the UAV is:
\begin{equation}
    R(s_k, a_k) = - r_{\text{mpn}} - \Vert x_k - x_{\text{goal}} \Vert_P^2
\end{equation}
where $r_{\text{mpn}}$ is the maximum $p$-norm of all $\Vert x(t) - x_{sp} \Vert_P^2$ evaluated during the entire SR cycle. The groundtruth state when the SR cycle has been completed is indicated by $x_k$. At any point in time if the system was becoming unstable ($\Vert x(t) - x_{sp} \Vert_P^2 >10$) we terminated the simulation with $r_{\text{mpn}}=10$.

\textbf{Baseline Method:} This method refers to the setpoint update \eqref{update_formula}. To make a fair comparison with the RL method, we used the following computation
\begin{equation} \label{eqn:alpha_contr}
    \alpha (\hat{x},x_{sp}) = \sqrt{\rho_m} - \Vert \hat x(t)-x_{sp} \Vert_p 
\end{equation}
where $\hat x(t)$ is the current state estimation, $x_{sp}$ denotes the current setpoint, and $\rho_m$ is the size of the outer ellipsoid set. The new formulation \eqref{eqn:alpha_contr} provides less conservative values than \eqref{update_formula}, because of \eqref{setpoint_condition} $\Vert \hat x(t)-x_{sp} \Vert_p \leq \sqrt{\rho_s}.$

\textbf{Reinforcement Learning Method:}
The RL method adopts the Soft Actor-Critic algorithm, with both Actor and Critic having two fully connected hidden layers 256 hidden units each. The model takes the difference between the state estimation and the current setpoint $\hat x(t) - x_{sp}$ as input and learns the optimal $\alpha$ value to generate the next setpoint under the safety constraints. We train the model with 20,000 steps on an NVIDIA GeForce RTX 3090, and the training takes approximately an hour to converge. During training, the policy is stochastic whereas during evaluation is deterministic.

\subsection{Simulation Results}

\begin{figure}[!htb]
    \centering
    \includegraphics[width=0.49\textwidth, trim={2.5cm 0cm 2cm 0cm},clip]{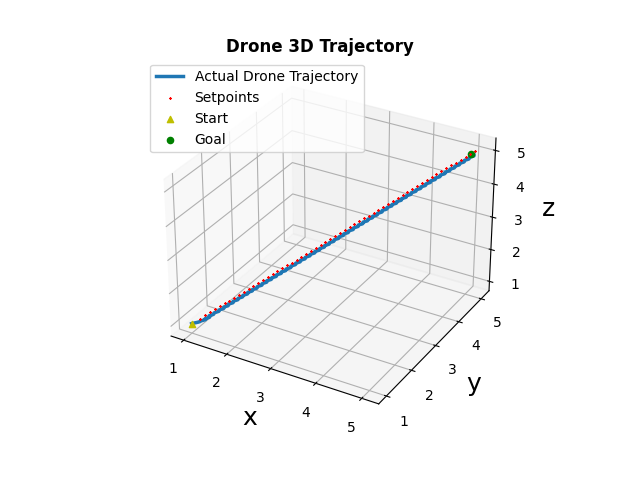}
    \caption{Drone trajectory with the baseline method. The drone starts from (1,1,1) and ends at (5,5,5). The red dots are the waypoints produced by the baseline method. The blue line shows the actual trajectory of the drone.}
    \label{fig:3d_traj_control}
\end{figure}

Under the baseline method, the drone completes the task in around 116s, and the drone's 3D trajectory is shown in Fig. \ref{fig:3d_traj_control}. The baseline method produces $\alpha$ based on equation  \eqref{eqn:alpha_contr} and generates setpoints shown as red dots in the figure. Our objective is to minimize the time required to reach the goal while ensuring that the system satisfies the safety conditions.

\begin{figure}[!htb]
    \centering
    \includegraphics[width=0.49\textwidth]{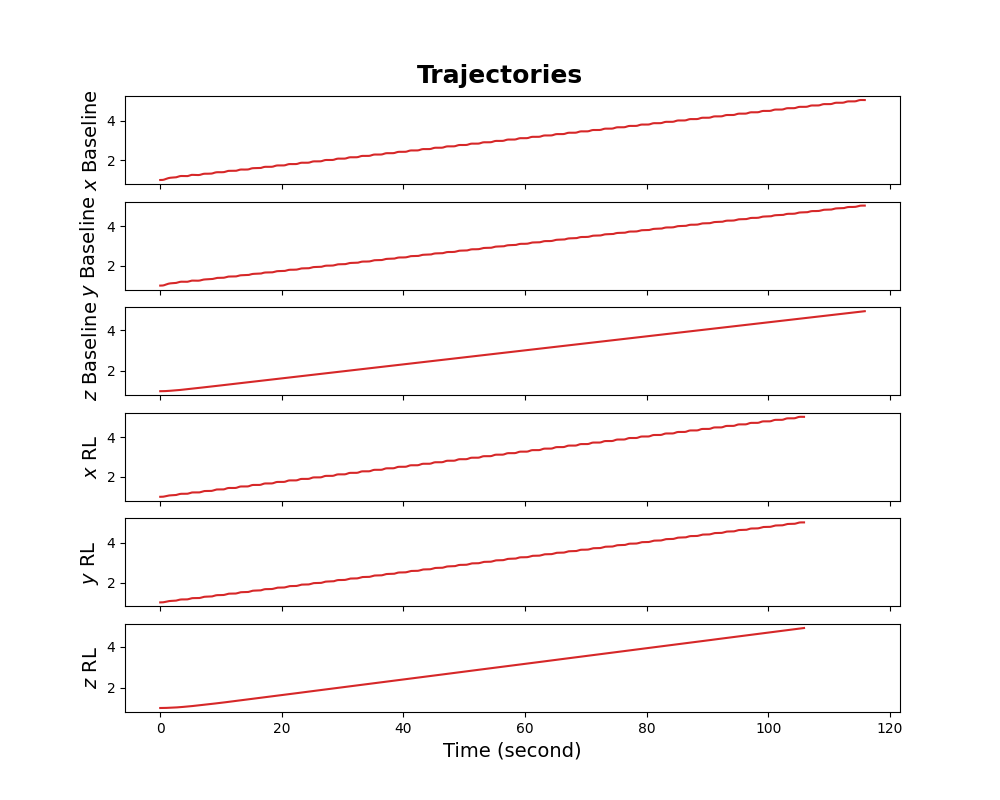}
    \caption{Drone trajectories of baseline method (top three plots) and RL method (bottom three plots). The baseline method takes 116 seconds, and the RL method takes 106 seconds to complete the same task.}
    \label{fig:state}
\end{figure}

\begin{figure}[!htb]
    \centering
    \includegraphics[width=0.49\textwidth]{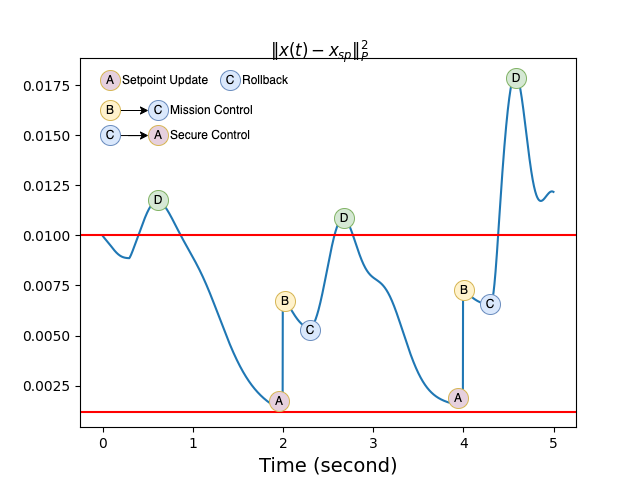}
    \caption{Example of $p$-norm analysis of the actual state w.r.t the current $x_{sp}$ selected from the first 5 s of our simulation with the baseline method. The red lines indicate the safety bounds $\rho_m=0.01$ and $\rho_s=0.0012$. (A, B, C, D, A) is a full checkpoint update cycle in Fig. \ref{fig:ES_switching}.}
    \label{fig:V_example}
\end{figure}

Our RL method reduces the total time required to complete the task, achieving the goal within 106 s, as demonstrated in the last three plots of Fig. \ref{fig:state}. In order to explain the benefit of the RL approach we consider the behavior of $\Vert x(t)-x_{sp}\Vert_P^2$ over  time. This function shows how far the system is from the boundaries of the ellipsoids $\mathcal{E}(\rho_m,x_{sp})$ and $\mathcal{E}(\rho_s,x_{sp})$ that guarantee the safety of the system under SR. Fig. \ref{fig:V_example} shows the first 5 s of the baseline method along with the safety bounds.



The sequence (A, B, C, D, A) forms a complete SR cycle, as shown in Fig. \ref{fig:ES_switching}. At local minima A, the setpoint is updated, and the system enters the $MC$ mode from B to C. To ensure safety, B points should be less than $\rho_m$, so $x \in \mathcal{E}(\rho_m, x_{sp})$. At point C, the software refreshes to the same value as the previous cycle's point B. From C to A, the system is in $SC$ mode, and $x$ can be outside $\mathcal{E}(\rho_m,x_{sp})$ as in D, which should be kept small to avoid stability issues.



\begin{figure}[!htb]
    \centering
    \includegraphics[width=0.49\textwidth]{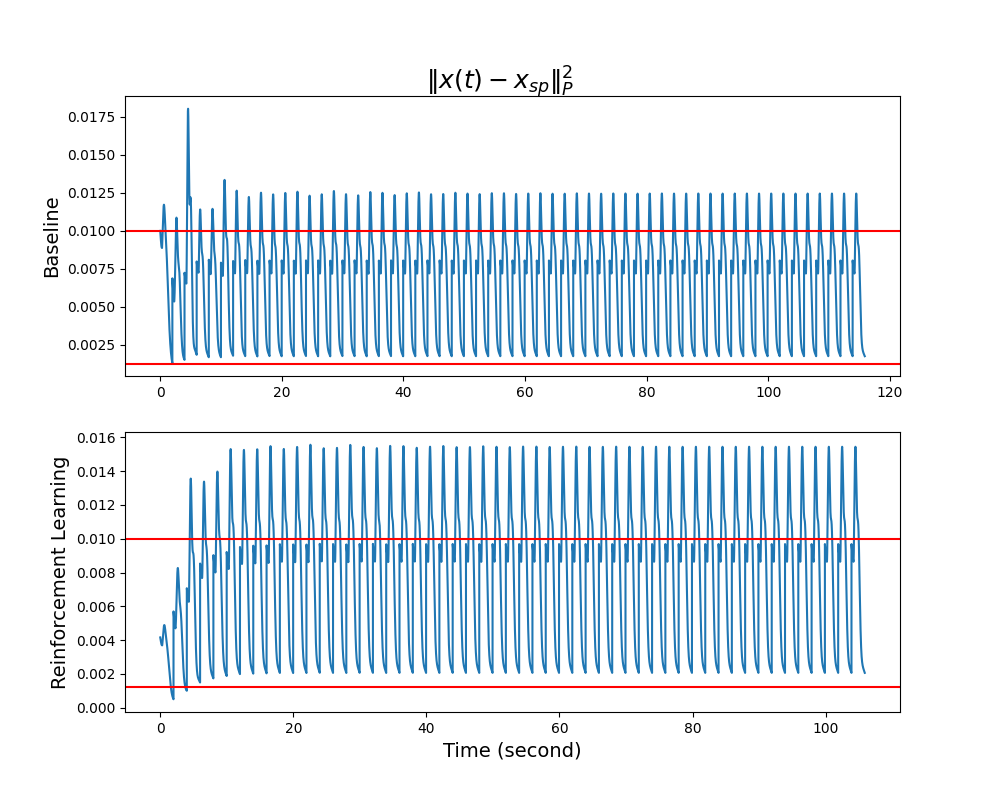}
    \caption{$p$-norm analysis of the actual state w.r.t the current $x_{sp}$. The top figure is $\Vert x(t)-x_{sp}\Vert_P^2$ with the baseline method, and the bottom one is for the RL method. The RL method pushes the equilibrium points higher and closer to the upper safety bound than the baseline method.}
    \label{fig:V_control}
\end{figure}

From Fig. \ref{fig:V_control}, we compare $\Vert x(t) - x_{sp} \Vert_P^2$ value between the baseline method and the RL method. The baseline method has B points at around 0.0075 during $MC$, while the RL method pushes B points up to 0.009, reducing the gap with the upper safety bound. 

\section{Conclusions}
This work demonstrated the effectiveness of incorporating Reinforcement Learning and optimal control methods in the design of safe and efficient UAV navigation systems. Our approach optimizes a reward function that balances safety and performance and incorporates Software Rejuvenation (SR) protection mechanisms to mitigate cyber-attacks. Results from simulations of UAVs show that our approach improves the system's performance while respecting the safety bounds compared to traditional methods. This work contributes to the growing body of research that combines control theory and Reinforcement Learning to address critical safety issues in cyber-physical systems. Future work can explore the application of our approach in other domains (e.g Bipedal walking) and investigate its impact on system performance and safety.

\section{Acknowledgements}
The authors would like to express their gratitude to Tao Jin and Prof. Anthony Rowe from the Wireless Sensing and Embedded Systems Lab at Carnegie Mellon University for their invaluable support in providing computational resources for this research.

\bibliographystyle{IEEEtran}
\bibliography{biblio}

\end{document}